\documentclass[]{aa}

\usepackage{graphicx}
\usepackage{natbib}

\begin{document}

\title{The hot stars in orbit around the M31 central supermassive black hole: are they young or old?}
\titlerunning{The hot stars near the M31 central black hole}
\authorrunning{P. Demarque \& S. Virani}

\author{Pierre Demarque \and Shanil Virani}
\institute{Department of Astronomy, Yale University, 
P.O. Box 208101, New Haven, CT 06520-8101}

\offprints{\email{pierre.demarque@yale.edu}, \email{shanil.virani@yale.edu}}


\date{Received: June 27, 2006 / Accepted: September 14, 2006}

\abstract
{}
{The cluster of hot stars observed in orbit around the central black 
hole of M31 has been interpreted 
as a 200 Myr starburst. The formation of a population of young stars 
in close proximity to a massive black hole presents a difficult 
challenge to star formation theory. We point out that in a high stellar density
environment, the course of stellar evolution is modified by 
frequent collisions and mergers. }
{Blue stragglers, which are the results of mergers in globular clusters, occupy the same position in the color-magnitude diagram as the observed hot stars in M31. For confirmation, the integrated spectrum of P3 is shown 
to be compatible with the spectral energy distribution of a blue horizontal branch field star.}  
{We suggest an old stellar population of 
evolved blue horizontal-branch stars and of merger products 
cannot be ruled out on the basis of
the available data. Observations are suggested that would help 
distinguish between a ``young'' and ``old'' stellar population 
interpretation of the observations. }    
{}


\maketitle

\section{Introduction}

Recent observations of the central region of M31 by \citet{bender05} have 
yielded convincing evidence for the presence of a central supermassive
black hole in M31.  
M31 is thus the third 
example of a galaxy with a well determined central black hole (BH) mass, after our 
Galaxy \citep{ghe03,ghe05,gen03}  
and M32 \citep{van98}.

\citet{bender05} identified three distinct stellar populations near the center of M31,
denoted as P1, P2 and P3.  P1 and P2 had been studied by previous authors \citep{lau93,kin95}.  
\citet{bender05} gave the name P3 to the compact source 
embedded within P2, whose strength in the ultra-violet had 
first been noted by \citet{kin95}.  These authors discovered  that the 
stellar population in P3 defined a 
disk like distribution orbiting the M31 central BH, and were thus able to 
derive a mass of the order of 1.1--2.3~ x $10^8$ $M_{\odot}$ for the massive BH, 
based on the measured velocities of the orbiting stars. 

From the spectrum, \citet{bender05},  
recognized the population substratum in P1 and P2 as an old stellar population.  
They identified the 
hot stars in P3, used in the central BH analysis, as young stars, and argue, with the 
help of stellar population models \citep{bru03,apa04}, that the spectrum 
is best interpreted as that of a 200 Myr starburst.

Although the dynamical argument used in deriving the mass of the  central BH 
is independent of the nature of the hot stars whose motions were 
measured, it is important to understand the evolutionary status of the stars 
in the central cusp of M31.
The identification of these stars as young stars poses challenging 
astrophysical problems regarding their origin, and about the star formation process 
in the close vicinity of a massive BH \citep{ale05}.  
It presents also dynamical difficulties if these stars were 
formed well away from the central BH and migrated towards it. The mediation of a  
second BH, of intermediate mass (in the range $10^3-10^4$ $M_{\odot}$) has 
been suggested \citep{han03}. 

In this Research Note, we address the issue of the stellar population and age of the 
observed hot stars. We point out 
that a case can be made that the observed 
spectrum does not necessarily belong to 
a 200 Myr starburst.  It could also belong to 
an old stellar population, in the post-HB evolutionary phase, or a population 
in which collisions 
have modified the course of stellar evolution.
While the young starburst hypothesis may well turn out to be correct, 
we note on the basis of the available information alone, that we 
cannot rule out the possibility that the observed stellar population is old.  The observed 
hot stars may be evolved stars that have lost 
most of their envelope, or that are the result of recent mergers in a dense environment.
The normal course of stellar evolution is modified in high stellar density environments 
\citep{ale05}.  The blue stragglers observed in globular star clusters are an 
example of such mergers 
\citep{bai95,sil99}.  Fig.~1, borrowed from \citet{sil99},  
illustrates the 
position of blue stragglers in the color-magnitude diagram.
The main point of this Research Note is to discuss this alternative 
interpretation, which if correct, would resolve some of the astrophysical difficulties 
associated with the young starburst hypothesis.  
The construction of models specifically applicable to M31
that take account of stellar interactions and mergers in high density regions 
is beyond the scope of this paper but will be the topic of a future paper.
Finally, we suggest observations 
that would help distinguish between the ``old'' and 
``young''  stellar population interpretations. 

This discussion would not be complete without 
recalling that a similar question can be asked on the formation process of the high 
velocity hot stars observed near the Galactic center, which 
are also usually believed to be very young.
It is tempting to draw an analogy between the Galactic center stars 
and those observed at the center of M31. We do this cautiously, however, fully aware 
that more detailed information is available on the 
environment of the hot stars observed in the Galactic center than in the case of M31.  
In addition, the Galactic center stars are B and O stars, much hotter 
than the A stars detected 
in M31, and it is possible that their origin may differ as well.  
 
\section{The discovery of hot stars in orbit around the central black hole in M31}

It is \citet{kin95} who discovered that P2 is much 
brighter than P1 in the ultra-violet.   They also pointed out that there is a compact source 
embedded in P2, which was later called P3 by \citet{bender05}. \citet{kin95}
also  noted that this 
compact source is similar in color and brightness to a 
single post-asymptotic giant branch (PAGB) star.  PAGB stars are observed in 
relatively old metal-rich stellar populations of the kind found in 
elliptical galaxies and galactic bulges. 
More than a decade ago, \citet{kin92} had found evidence from HST FOC data 
in the M31 bulge that half of the UV light came from PAGB stars. They attributed  
the rest of the observed UV background to evolved stars from a yet unresolved 
stellar population, possibly an old metal-rich population of the kind discussed 
by \citet{hor92} and \citet{cas92}.     
In a later study, \citet{kin95}   
favored a nonthermal light (AGN) interpretation, for the compact source in P2.

Further observations by \citet{lau98} and \citet{bro98}  
ruled out the single PAGB explanation by resolving the 
strong UV source into a cluster of stars.   
Finally, combining the observed ultra-violet  
and optical fluxes, \citet{lau98} concluded that 
the integrated light was consistent with a spectral energy distribution similar to   
that of an A star. 

The A star interpretation was recently confirmed by \citet{bender05},
who obtained the first integrated spectrum for the P3 peak embedded 
within P2 using STIS.
These authors demonstrated that the spectrum  
matches quite closely the spectrum of an A0 giant or A0 dwarf. After 
subtracting the background light from the P2 stellar population, they 
then made a plausible case that the stars in P3 are a 200 Myr starburst.  Although they were 
aware of the astrophysical difficulties posed by the existence of young stars so close 
to the M31 central BH, they merely commented that this phenomenon seems to occur in nature, 
appealing to the presence of hot stars near the    
Galactic central BH. The challenges to star formation theory and stellar dynamics 
near a massive BH 
have already given rise to an extensive literature in the context of the Galactic 
central BH  
\citep[see][ and references therein, for additional discussion]{nay05}.

\section{Evidence In Favor Of The Young Stellar Population Interpretation}

\subsection{Modeling the stellar population in P3}


\citet{bender05} have recently obtained the SED for P3 which
matches a dwarf or giant A0 stellar spectrum.  In
fact, for all practical purposes, a clump of stars in the
spectral range B5-A5, with
an integrated magnitude $M_v = -5.7$, would reproduce the spectrum
satisfactorily.  \citet{bender05} have interpreted their
observations as evidence for a starburst, which they have modeled using
the web based population synthesis program of \citet{bru03}.
The result favors an age of about 200 Myr, although the red end of the
spectrum matches a somewhat older population, with age of 510 Myr.
For illustration, they constructed
a CMD for a 200 Myr stellar population of solar metallicity using
the IAC-STAR web based population program \citep{apa04}.
\citet{bender05} scaled the model so that 186 main sequence stars
on the spectral range B5-A5 dominate the integrated spectrum.
Other parameters of the starbursts (IMF slope and mass range) are not
specified; they are indeed irrelevant since only the B5-A5 stars are
observable and the rest of the stellar
population is not detectable, because it is either too faint or because it
cannot be separated from the P2 stellar background.
The IAC-STAR model is confirmed by synthetic CMDs constructed for
a range of parameters using the web based population $Y^2$ synthesis program\footnote{http://www.astro.yale.edu/demarque/yyiso.html} which is based
on an independent grid of evolutionary tracks \citep{dem04}.
As emphasized above,
any population model
dominated by the light of A0 stars,
and whose integrated magnitude $M_v = -5.7$,
satisfies the observational constraint.

\subsubsection{A population of evolved stars}
Two comments must be made about the web based population models quoted above:
(1) They do not include, or include only in a approximate
way the advanced phases of evolution
for older stars which are known to produce hot stars in the CMD;
(2) Most importantly in the context of M31*, they describe
systems of non-interacting stars.  We know that
interactions between stars modify stellar evolution for many stars
in the central regions of globular clusters \citep{sil99,sil02}.
Since in the vicinity of the central BH, the stellar densities
can be two orders of magnitude higher than in a typical globular cluster
central region \citep{ale05}, interactions between individual stars
are extremely frequent.

In a realistic population modeling for M31*,
these two points must be taken into account.
More parameters must be considered in the construction
of synthetic models, such as the stellar densities and probabilities for
envelope stripping and stellar mergers.
Other key parameters including chemical
enrichment (the ratio $\Delta Y/\Delta Z$), amount of mass
transferred and heavy element content
must also be considered, making the task of population synthesis
more complex than simply producing a single CMD for non-interacting stars.
In addition, the products of stellar mergers in collisions
must be considered.

Relevant detailed stellar structure calculations and
population models already exist, well documented in the
literature, which justify the
plausibility of our argument (e.g. \citet{sil02,yi97b}).
For the present purpose, specialized studies
can be referred to, in particular Figures~7 to 9
in \citet{yi97b}.
The construction of models specifically applicable to M31
that take account of these effects is beyond the
scope of this paper but will be the topic of a future paper.

\section{Evidence in Favor For An Old Stellar Population Interpretation}

\subsection {Old hot stars and the UV upturn in elliptical galaxies}  

It has now become generally accepted that the UV upturn in elliptical galaxies 
is due to blue horizontal-branch (BHB) and post-HB stars \citep{yi99}.  
These stars are found in old stellar systems. Although the UV upturn is
attributed to the presence of B-type or hotter stars, only a modest variation in  
envelope mass and chemical composition is needed to create an HB 
morphology dominated by the light from A stars \citep{lee94}.  It was 
originally argued that 
high metallicity (likely in galaxy bulges) favor the production of luminous blue stars 
\citep{hor92,bressan94}, but further work has shown that it is 
not necessarily the case \citep{dor95,dcr96,yi97a,yi99}, 
and that several combinations of helium abundance 
and envelope mass could produce luminous blue stars.
A recent review by \citet{yi04} discusses 
the present status of the BHB interpretation for the UV upturn.  
Much better data have recently been obtained with the GALEX mission \citep[e.g][]{rey05};  
but there remains some uncertainties regarding the 
relative importance of metallicity, and the frequency of binary stars.  
In addition, as discussed 
below in this paper, the stripping of giant envelopes, and in some cases, stellar mergers 
could play an important role in regions of high stellar densities 
\citep{bai95,sil99,ale99}.

\citet{lee94} showed that the difference in morphology between a HB whose 
integrated light is dominated by B and O stars and another HB whose integrated 
light is dominated by A stars depends solely on plausible differences in 
chemical composition, age and envelope mass loss. 
We do not know the chemical composition nor 
age of the P3 stars in M31.  But we do have observational evidence for the existence of a population of evolved old 
stars in the bulge of M31, including luminous blue stars.  

In this paper, we propose that, within the uncertainties of current stellar 
evolution theory, the interpretation that P3 is made up of an evolved old stellar population, 
is a viable alternative to the young starburst interpretation.
Our point is illustrated in Fig.~2, where we have juxtaposed the observed spectrum \citet{bender05}  
with the spectrum of an AOV star and the spectrum of a field BHB star from the Sloan 
Digital Sky Survey \citep{sir04}.   

The observed spectrum could be either similar 
to the hot stars responsible for the UV upturn observed in elliptical galaxies and 
in the bulge of M31, or the result of close interactions due to high stellar densities \citep{ale99}.
Finding such stellar populations near the center of a galaxy, either as a result of past   
 star formation within the bulge, or through accretion of a dwarf satellite galaxy, would not be surprising 
in the context of what we know about the M31 bulge stellar populations and our  
current knowledge on galaxy evolution.

\subsection {Evidence from X-rays}

X-ray observations provide crucial additional information about the age
of the stellar population. X-rays would be expected to radiate from the
cool stars in a young starburst, due to their large magnetic activity. 
Young cool stars in star forming regions such as Orion have been
observed as X-ray sources.  The reported total hard X-ray luminosity
(2--8 keV) in the Orion nebula is 1.2$\times$10$^{33}$ erg s$^{-1}$
\citep{fei05}. Recently, \citet{gar05} reported a
3$\sigma$ upper limit to the emitted luminosity at the location of
M31$^*$ of 1.0 $\times$ 10$^{36}$ erg s$^{-1}$. This luminosity, 
however, is based only 13 net counts and clearly needs to be confirmed.
Nevertheless, even a confirmed X-ray detection of M31$^*$ does not
necessarily imply that the stars in its immediate vicinity are young,
as the detected X-rays are likely due to the presence of the AGN rather
than the X-ray emission from young stars as in Orion. Given the 
current spatial resolution of the Chandra X-ray Observatory, it is not
possible to separate the X-ray emission from a putative population of
young stars near M31$^*$ and from M31$^*$ itself (Generation-X should
be able to do so, however). On the other hand, if subsequent
observations do not detect X-ray emission from the center of M31, this
must mean that either the cool stars present are weak X-ray emitters
because the observed stellar population is old, or that there are too
few cool stars in the population, perhaps suggesting an IMF differing
from the canonical Salpeter law \citep{nay05}.

\subsection{Possible analogy with the center of our Galaxy}
Although the situation may be quite different in the Galaxy and in M31, we 
recall the similarities  
with the region of the central BH of our Galaxy.  This similarity was 
appealed to by \citet{bender05} to explain the plausibility of observing young stars near the central BH.  
We draw this parallel tentatively because there are a number 
of differences between the observations in 
the Galaxy and M31.  The high velocity hot stars observed in the Galaxy are B and O stars, 
markedly hotter than the A stars that dominate the M31 spectrum.  If they are old stars, 
because HB morphology is so sensitive to details of chemical composition and age, and to the 
binary fraction, in ways that are not fully understood, they need only 
differ in one or several of these characteristics.

Most importantly,  the central cusp environment favors the stripping of red giant envelopes and the 
occurrence of stellar mergers.  All these factors favor the formation of hot luminous stars in old 
stellar populations.    

\subsubsection{The hot stars in the Galactic center}

High velocity stars with early spectral types (O and B) have been observed 
near the center of the Galaxy, in close proximity to the  central BH \citep{ghe03,ghe05,gen03}.

 In the central region of the Galaxy, outside the cusp, \citet{gen03}  
concluded, from the K-band luminosity function that 
the stellar population was relatively old (about 8 Gyrs) and characterized by a red giant branch and 
a red clump of core helium burning stars.  This underlying stellar population is characteristic 
of the central bulge of the Galaxy.  \citet{gen03} then contrasted  the stellar K luminosity 
function within the central cusp.  They state that the ``cusp within $\leq$ 1.5\arcsec ~of Sgr A* appears 
to have a featureless luminosity function, suggesting that old, low mass, HB/red clump stars are lacking''.  
They further state ``likewise there appears to be fewer late-type giants.  The innermost cusp also 
contains a group of moderately bright, early-type stars that are tightly bound to the BH.''

The striking difference between the inner cusp is significant, and the underlying 
stellar population could be 
explained in several ways.  The favored interpretation of most has been that the BH environment 
somehow favors star formation, and in particular the formation of massive stars \citep{nay05}, 
although \citet{gen03} consider a possible merger interpretation for the OB stars 
within 0.04 pc of the Galactic center, named the S-stars by \citet{ghe98}.  
Another interpretation is that dynamical interactions in the old stellar 
population is entirely responsible for the cusp phenomenon.  The higher stellar densities cause close 
interactions between stars.  The frequency of envelope stripping on the red giant branch, 
and/or stellar mergers, would be enhanced.  Some of the 
stripped giants would never reach core helium burning, others would evolve as blue HB and post-HB 
objects.  In addition, the probability of mergers is enhanced.  All these lead to the 
formation of a variety of luminous blue 
stars in A or hotter temperature range.  
The entire stellar population would then be old, but characterized by a deficiency of red giants 
and of red HB stars, exactly what is observed.  

A final comment can be made about the spectral identification of individual stars 
in the Galactic center vicinity. 
We note that these objects are 
highly obscured and reddened by interstellar material.  Often referred to 
as ``HeI stars'' because of their helium emission or absorption features ,  their 
spectra have been studied by a number of authors \citep{kra95,tam96,blu96,eckart97,pau01}.
  The spectra 
show a variety of features, and can be divided into several classes.  \citet{pau01}'s discussion
highlights the diversity of their spectra, which have given rise to differences 
in identification and classification between independent teams of researchers. 
Since  most of the hot stars in a gas rich environment which have been studied 
spectroscopically in detail are  
young stars, the Galactic center spectra have been compared to those.  Young hot stars 
are massive stars, found in open star clusters or associations,  and accordingly, the 
observed ``HeI stars'' have been assigned large masses and young ages. 

Stellar spectra by themselves do not contain any information about the 
ages of stars.  
Stellar spectra depend sensitively on temperature and chemical
 composition, and to some extent on surface gravity.  In the absence of 
other information, such as physical environment, or the population to which they belong, 
 it is difficult to recognize a young star
from an old star simply on the basis of its temperature and luminosity.
The identifications made so far were made simply by analogy with young starbursts 
that have been 
observed in other regions of the Galaxy, well removed from the Galactic center.  

 It is well 
recognized that if these stars are young, their origin is 
difficult to explain.  In addition, the lack of X-rays from the Galactic center
can only be 
explained by introducing a non-standard IMF \citep{sal55,mil79}  
as recently suggested by \citet{nay05}.
Before accepting this important implication for star formation, one 
needs also to definitively rule out the possibility that the 
hot stars observed near the Galactic central BH belong to an old 
stellar population.  

\subsection{Dynamical studies}
The disruption of a captured massive old star cluster or dwarf 
galaxy into a disk gravitating 
about the central BH has been proposed as an interpretation. Monte Carlo 
simulations applied to a dense star cluster near the center of our own Galaxy 
\citep{gur05} seem compatible with such an interpretation, 
as long as the cluster is sufficiently massive ($10^6$ $M_{\odot}$).  
We note, however, that the dynamical models of \citet{gur05}, 
designed for young star clusters, would not be 
directly applicable in the case of the capture of an old group of stars. 
  In a young system, a large range of masses along the main sequence coexist, with 
the largest masses up to about 50$ M_{\odot}$.  By contrast, the most luminous 
stars in the UV  in an old stellar system have progenitors with 
masses of the order of one solar mass.  
By the time they have evolved to the blue horizontal branch (BHB) and beyond, stars have lost about half 
of their original mass.  In the case of stellar mergers, the merger product would 
approximately double its mass.  
This fundamental difference in mass range 
would modify the gravitational segregation 
and selective evaporation in the infalling cluster.

The central 0.1 pc around the central BH are characterized by high 
stellar densities (up to about $10^8$ $M_{\odot}$ per cubic parsec), 
or two orders of magnitude higher than at the center of globular 
star clusters.  Stellar collisions are therefore very frequent.
Two-body interactions lead to star segregation with the more massive 
stars being more centrally concentrated.

Observations indicate that the old stars are well relaxed in their velocity 
distribution while the 
hot stars are not. Separating the two stellar populations 
is the main problem in interpreting the velocity data. A population of 
old stars which have relatively recently lost much of their mass 
due to mass loss and stellar interactions (as is the case for old 
blue stars), would not be dynamically relaxed.  Some 
stars could 
be kicked up to much higher velocities than their 
progenitors, particularly in binary systems.

The dynamical interaction would lead to an excess of randomly distributed 
high velocities, which is observed among the S-stars.  Other evidence 
for unusually high velocities due to direct interaction with the BH is 
provided by the recent discovery of hypervelocity stars (HVS),
described in the next section.

We have pointed out that the bulge of our Galaxy and M31 are known to be 
primarily populated with old stars. In addition, 
the capture of old stellar systems by the Galactic halo has been extensively 
documented \citep{iba03,duf06}.  Finding 
the remnants of captured old stellar populations   
in the inner part of the galactic bulge would not be surprising.  
For example, a concentration of A stars is expected in HB models for Z = 0.01 
and an age of several Gyrs (e.g. see 
Fig.~8 in \citealt{yi97b}).  Even without accounting for mass loss 
due to tidal interactions, this would not be 
unreasonable for a typical satellite dwarf galaxy population 
near a large spiral galaxy like M31.    

\subsubsection{Hypervelocity stars}

The plausibility of finding hot old stars near the Galactic central BH is 
shown by the recent discovery of hypervelocity stars (HVSs) in the Galaxy.
Their existence lends support to the conclusion that at 
least some of the B stars in the vicinity of the Galactic center are old. 
HVSs are believed to achieve their extremely high velocities 
by dynamical ejection from a binary system while interacting with 
a massive black hole \citep{hil88}.  Five HVSs have so far 
been discovered \citep{bro05,bro06,hir05,ede05}.  
If Hills's interpretation is correct, hypervelocity stars 
provide some information on the stellar population in the vicinity of the 
massive black hole in the Galaxy. We note that two of the discovered HVSs 
\citep{hir05,ede05} have been identified as 
likely blue HB (sdB) candidates.
The other three are also B stars observed by \citet{bro05,bro06}.
They were discovered in a search for blue HB stars and could be either main 
sequence B stars or subdwarf B stars (sdB). We know that sdBs 
evolve into luminous hot stars \citep{yi99}.

\section {Suggested observational tests}
A crucial test of the ``old stars'' interpretation for the stellar population near the 
M31 central BH could be made by 
detecting the presence of the less luminous red giants and main sequence stars 
that are the signature of an old stellar system.  Spectral information in 
the red is needed.  In particular, spectroscopic evidence for 
the presence of late spectral type giants and dwarfs in P3, 
or the lack thereof, 
would provide a crucial test 
of the old vs. young interpretation.
At the present time, X-ray observations of M31 are 
insufficiently sensitive, and lack the required spatial resolution to 
make any statement about the presence (or absence) in P3
of young magnetically active cool stars, but future space missions will
make the observations possible. 
Another test of the presence of young stars would be the detection of 
infrared excesses due to 
the presence of circumstellar disks \citep{yud00}. 

In our Galaxy, ongoing searches for HVSs, and more detailed studies of their 
characteristics will provide further indications of the stellar 
population near the massive BH. At this point, the available data from the 
HVSs,
admittedly very scant,  
show that at least a fraction of the stars close to the 
massive BH at the Galactic center are old HB stars.

\section{Summary}

In this paper, we propose that the hot stars observed near the central
BH of M31 could belong to an old stellar population, similar to the 
stellar population observed in the M31 bulge.  The unusual 
physical environment  (high stellar densities) in the vicinity 
of the central BH is expected to modify the normal course of stellar 
evolution in several possible ways:
\begin{enumerate}
\item By the stripping of a part or of the whole envelope of a red giant 
envelope.  The end result is the formation of a blue HB star, either a sdB 
or a sdO, depending on the amount of mass loss, rather that red clump stars.  
Further, blue HB stars 
evolve into hot luminous stars instead of becoming asymptotic giants.
\item Through stellar collisions resulting in mergers. Such merger products 
form a population of luminous hot stars analogous to the blue stragglers 
observed in globular star clusters. 
\end{enumerate}

An old stellar population in which collisions are frequent would be 
characterized by a defect of red horizontal branch stars (red clump 
stars) and luminous red giants when compared to a normal population 
of the same age and chemical composition.  At the same time, it should
exhibit a population of hot stars of various luminosities and colors.
In addition, having suffered recent collisions, the space motions of the 
hot star population would differ from that of the background population, 
and in particular could exhibit unusually high space velocities. 
The high velocity S-stars  
observed within    pc of the Galactic central BH may be examples 
of such stars.  
In the extreme 
case of binary systems interacting with the central BH itself, 
close interaction could give rise to hypervelocity stars. 

The old star interpretation has several advantages over the young 
stellar population interpretation.  It removes the problem of 
understanding star forming regions $\emph{in situ}$ near the 
massive BH in the center of M31 and the Galaxy.  In the case of the Galaxy,
it explains naturally the absence of strong X-ray sources without 
appealing to a non-standard IMF \citep{nay05}. Moreover,
\citet{jimenez06} in a recent preprint suggest that irradiation of a stellar
atmosphere by an AGN can make late type stars appear young. Therefore, the
confluence of these several independent lines of investigation suggest that
the stars in the immediate vicinity of M31* could be explained by a population
of old stars.

\acknowledgements
We thank the referee for comments that have improved the paper. 
The authors are indebted to P. Coppi, R.B. Larson, T. Lauer, E. Murphy, 
J.-H. Woo and 
R.J. Zinn for helpful conversations on the stellar populations near the 
central BHs of M31 and the Galaxy.  We have also benefitted from useful 
correspondence from W. Brown, T. Kinman and T. Davidge. Special thanks 
are due to P. Natarajan for reading an early draft of the paper and for 
suggesting helpful revisions that have improved the manuscript.    

This work was supported in part by NASA grant NAG5-13299.     

\bibliographystyle{aa}

\begin{figure*}
\centering
\includegraphics[width=15.25cm,angle=90]{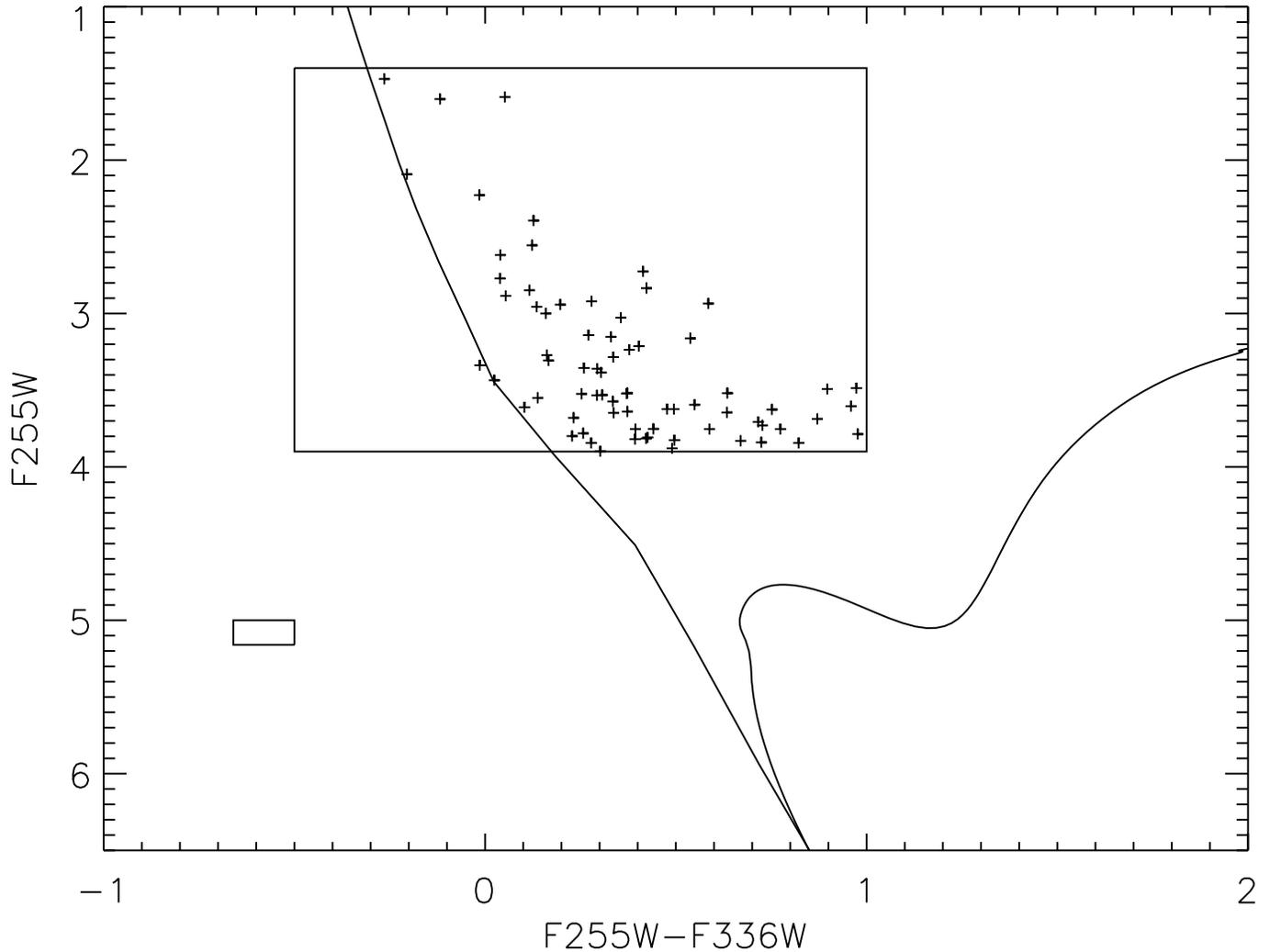}
\caption{Schematic color-magnitude diagram of a globular star cluster in the 
HST ultraviolet filters F255W and F336W. 
The box indicates the position of 
observed blue straggler stars, believed to be the results of mergers.  The 
position of the zero-age main sequence is marked by a solid line. The other 
solid line is the position of the fiducial stellar sequence of the cluster. 
The plotted blue straggler data are from \citet{fer97}. This figure is taken 
from \citet[][ Fig. 3]{sil99}.}
\label{fig1}
\end{figure*}

\begin{figure*}
\centering
\includegraphics[width=16.25cm,angle=0]{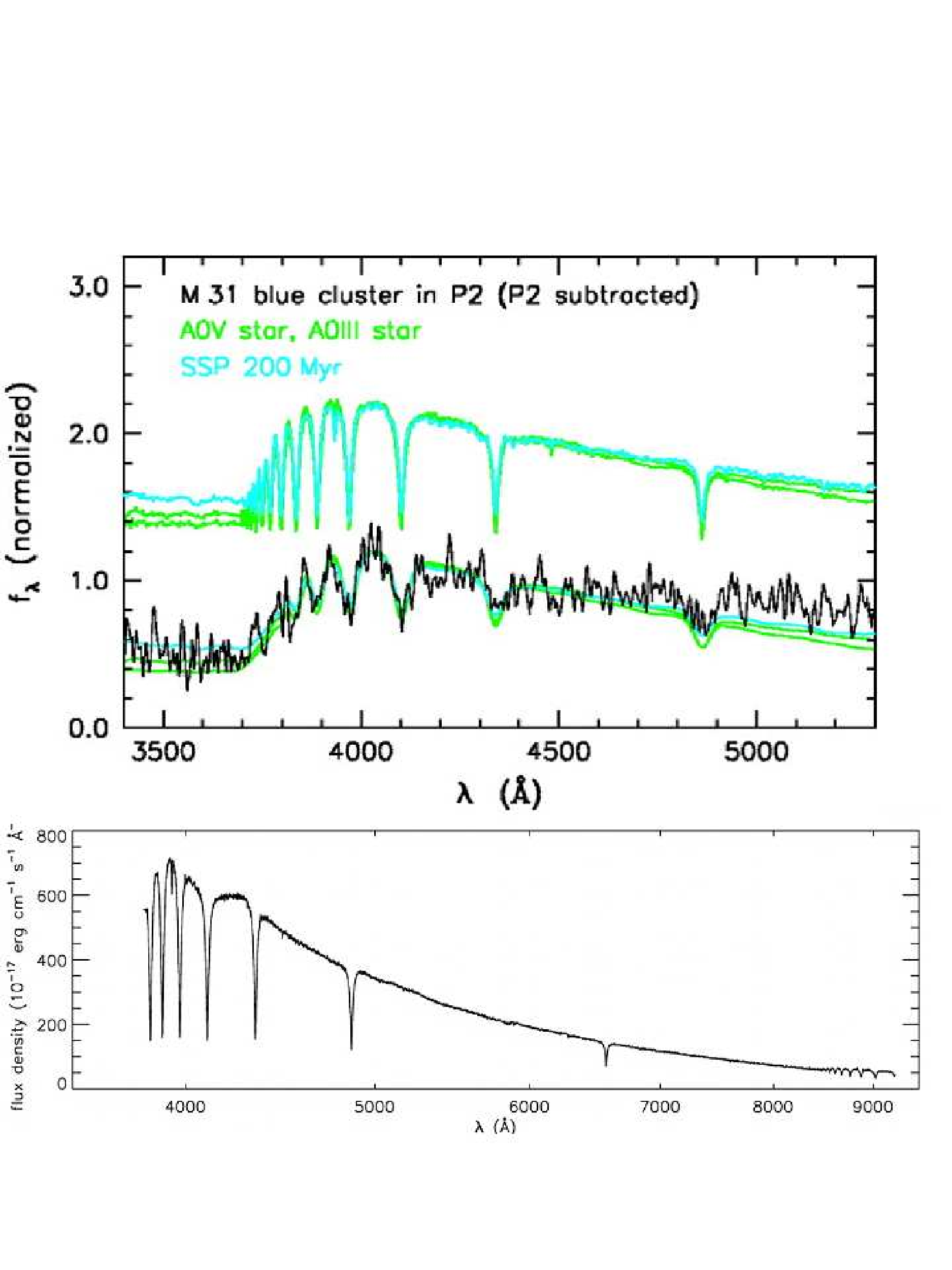}
\caption{The observed SED of P3 in M31 \citep{bender05} is the black curve 
shown in the top panel.  
It is compared to the SEDs of 
an A0 dwarf and giant, above (adapted from \citealp{bender05}, Fig.4), 
and to the SED of a blue HB field 
star in the Galactic halo, below (from Fig. 2 of \citealp{sir04}). 
At the available spectral 
resolution, either stellar SED could fit the observed SED.}
\label{fig2}
\end{figure*}

\end{document}